\pdfoutput=1
\documentclass[article]{aa}
\usepackage{graphicx}
\usepackage{txfonts}
\usepackage[]{hyperref}
\usepackage{longtable}
\graphicspath{{./images/}}
\bibpunct{(}{)}{;}{a}{}{,} 

\begin{document}

   \title{The Solar Twin Planet Search}

   \subtitle{IV. The Sun as a typical rotator and evidence for a new \\rotational braking law for Sun-like stars\thanks{Based on observations collected at the European Organisation for Astronomical Research in the Southern Hemisphere under ESO programs 188.C-0265, 183.D-0729, 292.C-5004, 077.C-0364, 072.C-0488, 092.C-0721, 093.C-0409, 183.C-0972, 192.C-0852, 091.C-0936, 089.C-0732, 091.C-0034, 076.C-0155, 185.D-0056, 074.C-0364, 075.C-0332, 089.C-0415, 60.A-9036, 075.C-0202, 192.C-0224, 090.C-0421 and 088.C-0323.}}

   \author{Leonardo A.~dos Santos
          \inst{1,2}
          \and
          Jorge~Mel\'{e}ndez\inst{1}
          \and
          Jos\'{e}-Dias~do Nascimento Jr.\inst{3,4}
          \and
          Megan~Bedell\inst{2}
          \and
          Iv\'an~Ram{\'{\i}}rez\inst{5}
          \and
          Jacob L.~Bean\inst{2}
          \and
          Martin~Asplund\inst{6}
          \and
          Lorenzo~Spina\inst{1}
          \and
          Stefan~Dreizler\inst{7}
          \and
          Alan~Alves-Brito\inst{8}
          \and
          Luca~Casagrande\inst{6}
          }

   \institute{Universidade de S\~ao Paulo, Departamento de Astronomia do IAG/USP, Rua do Mat\~ao 1226,
             Cidade Universit\'aria, 05508-900 S\~ao Paulo, SP, Brazil \\
             \email{leonardoags@usp.br}
         \and
             University of Chicago, Department of Astronomy and Astrophysics, USA
         \and
             Universidade Federal do Rio Grande do Norte, 59072-970 Natal, RN, Brazil
         \and
             Harvard-Smithsonian Center for Astrophysics, Cambridge, MA 02138, USA
         \and
             University of Texas, McDonald Observatory and Department of Astronomy at Austin, USA
         \and
             The Australian National University, Research School of Astronomy and Astrophysics, Cotter Road, Weston, ACT 2611, Australia
          \and
             University of G\"ottingen, Institut f\"ur Astrophysik, Germany
          \and
             Universidade Federal do Rio Grande do Sul, Instituto de F\'isica, Av. Bento Gon\c{c}alves 9500, Porto Alegre, RS, Brazil
          }

   \date{Received 19 March 2016; accepted 20 June 2016}

  \abstract
   {It is still unclear how common the Sun is when compared to other similar stars in regards to some of its physical properties, such as rotation. Considering that gyrochronology relations are widely used today to estimate ages of stars in the main sequence, and that the Sun is used to calibrate it, it is crucial to assess if these procedures are acceptable.}
   {We analyze the rotational velocities -- limited by the unknown rotation axis inclination angle -- of an unprecedented large sample of solar twins in order to study the rotational evolution of Sun-like stars, and assess if the Sun is a typical rotator.}
   {We use high-resolution ($R = 115000$) spectra obtained with the HARPS spectrograph and ESO's 3.6 m telescope at La Silla Observatory. The projected rotational velocities for 82 solar twins are estimated by line profile fitting with synthetic spectra. Macroturbulence velocities are inferred from a prescription that accurately reflects their dependence with effective temperature and luminosity of the stars.}
   {Our sample of solar twins include some spectroscopic binaries with enhanced rotational velocities, and we do not find any non-spectroscopic binaries with unusually high rotation velocities. We verified that the Sun does not have a peculiar rotation, but the solar twins exhibit rotational velocities that depart from the Skumanich relation.}
   {The Sun is a regular rotator when compared to solar twins with a similar age. Additionally, we obtain a rotational braking law that better describes the stars in our sample ($v \propto t^{-0.6}$) in contrast to previous, often-used scalings.}

   \keywords{Sun: rotation --
             stars: solar-type --
             stars: rotation --
             stars: fundamental parameters
               }

   \maketitle


\section{Introduction}

The Sun is the best known star to astronomers, and is commonly used as a template in the study of other similar objects. Yet, there are still some of its aspects that are not well understood and that are crucial for a better understanding of how stars, and consequently how planetary systems and life evolve: how do the more complex physical parameters of a Sun-like star, such as rotation and magnetic activity, change with time? Is the Sun unique or typical (i.e., an average Sun-like star)? If the Sun is common, it would mean that life does not require a special star for it to flourish, eliminating the need to evoke an anthropic reasoning to explain it.

In an effort to assess how typical the Sun is, \citet{2008ApJ...684..691R} compared 11 of its physical parameters with nearby stars, and concluded that the Sun is, in general, typical. Although they found it to be a slow-rotator against 276 F8 -- K2 (within $\pm 0.1$ M$_\odot$) nearby stars, this result may be rendered inconclusive owing to unnacounted noise caused by different masses and ages in their sample. Other studies have suggested that the Sun rotates either unusually slow \citep{1979PASP...91..737S, 2015A&A...582A..85L} or regularly for its age \citep{1983ApJS...53....1S, 1985AJ.....90.2103S, 1984ApJ...281..719G, 1998SSRv...85..419G, 2003ApJ...586..464B}, but none of them comprised stars that are very similar to the Sun, therefore preventing a reliable comparison. In fact, with \textit{Kepler} and \textit{CoRoT}, it is now possible to obtain precise measurements of rotation periods, masses and ages of stars in a very homogeneous way \citep[e.g.,][]{2015EPJWC.10106016C, 2012A&A...548L...1D, 2014ApJS..210....1C}, but they generally lack high precision stellar parameters, which are accessible through spectroscopy. The challenging nature of these observations limited ground-based efforts to smaller, but key stellar samples \citep[e.g.,][]{2003A&A...397..147P, 2012AN....333..663S}.

The rotational evolution of a star plays a crucial role in stellar interior physics and habitability. Previous studies proposed that rotation can produce extra mixing that is responsible for depleting the light elements Li and Be in their atmospheres \citep{1989ApJ...338..424P, 1994A&A...283..155C, 2015A&A...576L..10T}, which could explain the disconnection between meteoritic and solar abundances of Li \citep{2010A&A...519A..87B}. Moreover, rotation is highly correlated with magnetic activity \citep[e.g.,][]{1984ApJ...279..763N, 1993ApJS...85..315S, 1995ApJ...438..269B, 2008ApJ...687.1264M}, and this trend is key to understand how planetary systems and life evolve in face of varying magnetic activity and energy outputs by solar-like stars during the main sequence \citep{2009IAUS..258..395G, 2005ApJ...622..680R, 2016ApJ...820L..15D}.

A theoretical treatment of rotational evolution from first principles is missing, so we often rely on empirical studies to infer about it. One of the pioneer efforts in this endeavor produced the well known Skumanich relation $v \propto t^{-1/2}$, where $v$ is the rotational velocity and $t$ is the stellar age \citep{1972ApJ...171..565S}, which describes the rotational evolution of solar-type stars in the main sequence, and can be derived from the loss of angular momentum due to magnetized stellar winds \citep[e.g.,][]{1988ApJ...333..236K, 1992ASPC...26..416C, 2003ApJ...586..464B, 2013A&A...556A..36G}. This relation sparked the development of gyrochronology, which consists in estimating stellar ages based on their rotation, and it was shown to provide a stellar clock as good as chromospheric ages \citep{2007ApJ...669.1167B}. However, in Skumanich-like relations, the Sun generally falls on the curve (or plane, if we consider dependence on mass) defined by the rotational braking law by design. Thus it is of utmost importance to assess how common the Sun is in order to correctly calibrate it.

Subsequent studies have proposed modifications to this paradigm of rotation and chromospheric activity evolution \citep[e.g.,][]{1991ApJ...375..722S, 2004A&A...426.1021P}, exploring rotational braking laws of the form $v \propto t^{-b}$. The formalism by \citet{1988ApJ...333..236K} shows that this index $b$ can be related to the geometry of the stellar magnetic field, and that Skumanich's index ($b = 1/2$) corresponds to a geometry that is slightly more complex than a simple radial field. It also dictates the dependence of the angular momentum on the rotation rate, and in practice, it determines how early the effects of braking are felt by a model. Such prescriptions for rotational evolution have a general agreement for young ages up to the solar age \citep[see][and references therein]{2016arXiv160507125S, 2016A&A...587A.105A}, but the evolution for older ages still poses an open question. In particular, \citet{2016Natur.529..181V} suggested that stars undergo a weakened magnetic braking after they reach a critical value of the Rossby number, thus explaining the stagnation trend observed on the rotational periods of older Kepler stars.

\begin{figure}[!t]
   \centering
   \includegraphics[width=\hsize]{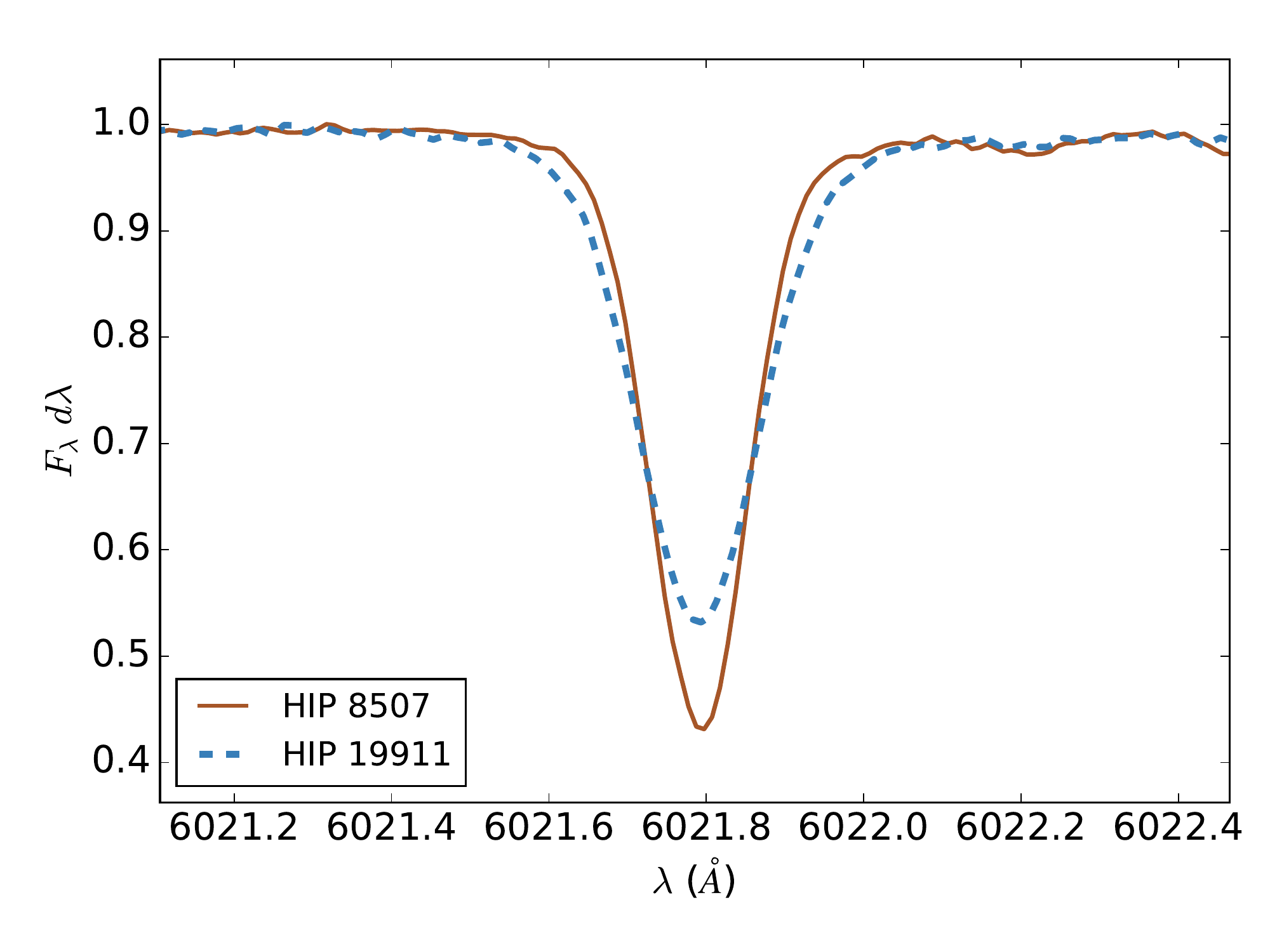}
   \caption{Comparison of the spectral line broadening between two solar twins with different projected rotational velocities. The wider line correspond to HIP 19911, with $v \sin i \approx 4.1$ km s$^{-1}$, and the narrower one comes from HIP 8507, with $v \sin i \approx 0.8$ km s$^{-1}$.}
   \label{widths}
   \end{figure}

In order to assess how typical the Sun is in its rotation, our study aims to verify if it follows the rotational evolution of stars that are very similar to it, an objective that is achieved by precisely measuring their rotational velocities and ages. We take advantadge of an unprecedented large sample of solar twins \citep{2014A&A...572A..48R} using high signal-to-noise ($S/N > 500$) and high resolution ($R > 10^5$) spectra, which provides us with precise stellar parameters and is essential for the analysis that we perform (see Fig. \ref{widths} for an illustration of the subtle effects of rotation in stellar spectra of Sun-like stars).


\section{Working sample}

Our sample consists of bright solar twins in the Southern Hemisphere, which were mostly observed in our HARPS Large Program (ID: 188.C-0265) that aimed to search for planetary systems around stars very similar to the Sun \citep[][Papers I, II and III, respectively, of the series The Solar Twin Planet Search]{2014A&A...572A..48R, 2015A&A...581A..34B, 2016A&A...590A..32T}. These stars are loosely defined as those that have T$\mathrm{_{eff}}$, $\log{g}$ and [Fe/H] inside the intervals $\pm 100$ K, $\pm 0.1$ [cgs] and $0.1$ dex, respectively, around the solar values. It has been shown that these limits guarantee $\sim$0.01 dex precision in the relative abundances derived using standard model atmosphere methods amd that the systematic uncertainties of that analysis are negligible within those ranges \citep{2014ApJ...795...23B, 2015A&A...583A.135B, 2015A&A...582A..17S, 2016A&A...589A..17Y}. In total, we obtained high precision spectra for 73 stars and used data from 9 more targets observed in other programs, all of them overlapping the sample of 88 stars from Paper I. We used the spectrum of the Sun (reflected light from the Vesta asteroid) from the ESO program 088.C-0323, which was obtained with the same instrument and configuration as the solar twins.

The ages of the solar twin sample span between $0-10$ Gyr and are presented in the online material (Table \ref{params}). They were obtained by \citet{2016A&A...590A..32T} using Yonsei-Yale isochrones \citep{2001ApJS..136..417Y} and probability distribution functions as described in \citet{2013ApJ...764...78R,2014A&A...572A..48R}. Uncertainties are assumed to be symmetric. These ages are in excellent agreement with the ones obtained in Paper I, with a mean difference of $-0.1 \pm 0.2$ Gyr (see footnote 5 in Paper III). We adopted 4.56 Gyr for the solar age \citep{1995RvMP...67..781B}. The other stellar parameters ($T\rm_{eff}$, $\log{g}$, [Fe/H] and microturbulence velocities $v\rm_t$) were obtained by \citet{2014A&A...572A..48R}. The stellar parameters of HIP 68468 and HIP 108158 were updated by \citet{2016A&A...590A..32T}.

Our targets were observed at the HARPS spectrograph\footnote{\footnotesize{The initial plan was to use the observations from the MIKE spectrograph, as described by the Paper I. However, we decided to use the HARPS spectra due to its higher spectral resolving power.}} \citep{2003Msngr.114...20M} which is fed by ESO's 3.6 m telescope at La Silla Observatory. When available publicly, we also included all observations from other programs in our analysis in order to increase the signal to noise ratio ($S/N$) of our spectra. However, we did not use observations for 18 Sco (HIP 79672) from May 2009\footnote{\footnotesize{These observations have instrumental artifacts.}} and we did not include observations post-HARPS upgrade (June 2015) when combining the spectra\footnote{\footnotesize{The spectra had a different shape in the red side, and since there were few observations, we chose not to use them to eliminate eventual problems with combination and normalization.}}.

The wavelength coverage for the observations ranged from 3780 to 6910 \AA, with a spectral resolving power of $R = \lambda/\Delta\lambda = 115000$. Data reduction was performed automatically with the HARPS Data Reduction Software (DRS). Each spectrum was divided in two halves, corresponding to the mosaic of two detectors (one optimized for the blue and other for the red wavelengths). In this study we only worked with the red part (from 5330 to 6910 \AA) due to its higher $S/N$ and the presence of cleaner lines. The correction for radial velocities was performed with the task \texttt{dopcor} from IRAF\footnote{\footnotesize{IRAF is distributed by the National Optical Astronomy Observatories, which are operated by the Association of Universities for Research in Astronomy, Inc., under cooperative agreement with the National Science Foundation.}}, using the values obtained from the pipeline's cross-correlation function (CCF) data. The different observations were combined with IRAF's \texttt{scombine}. The resulting average (of the sample) signal to noise ratio was 500 around 6070 \AA. The red regions of the spectra were normalized with $\sim$30th order polynomial fits to the upper envelopes of the entire red range, using the task \texttt{continuum} on IRAF. We made sure that the continuum of the stars were consistent with the Sun's. Additionally, we verified that errors in the continuum determination introduce uncertainties in $v \sin{i}$ lower than $0.1$ km s$^{-1}$.

\section{Methods}\label{methods}

We analyze five spectral lines, four due to Fe I and one to Ni I (see Table \ref{lines}; equivalent widths were measured using the task \texttt{splot} in IRAF.), that were selected for having low level of contamination by blending lines. The rotational velocity of a star can be measured by estimating the spectral line broadening that is due to rotation. The rotation axes of the stars are randomly oriented, thus the spectroscopic measurements of rotational velocity are a function of the inclination angle ($v \sin{i}$).

\begin{table}[h]
\begin{center}
\caption{Line list used in the projected stellar rotation measurements.}
\begin{tabular}{cccccc}
\hline \hline\\[-2ex]
Wavelength & Z & Exc. pot. & $\log{(gf)}$ & $v_{\mathrm{macro}}^\odot$ & EW$^\odot$\\
(\AA) &  & (eV) & & (km s$^{-1}$) & (\AA)\\ \hline
6027.050 & 26 & 4.076 & -1.09  & 3.0 &	0.064 \\
6151.618 & 26 & 2.176 & -3.30 & 3.2 &	0.051 \\
6165.360 & 26 & 4.143 & -1.46  & 3.1 &	0.045 \\
6705.102 & 26 & 4.607  & -0.98  & 3.6 &	0.047 \\
6767.772 & 28 & 1.826 & -2.17  & 2.9 &	0.079 \\
\hline
\end{tabular}
\tablefoot{EW are the equivalent widths and $v_{\mathrm{macro}}$ are the macroturbulence velocities measured as in Sect. \ref{vmacro_det}.}
\label{lines}
\end{center}
\end{table}

\begin{figure}[!t]
\centering
\includegraphics[width=\hsize]{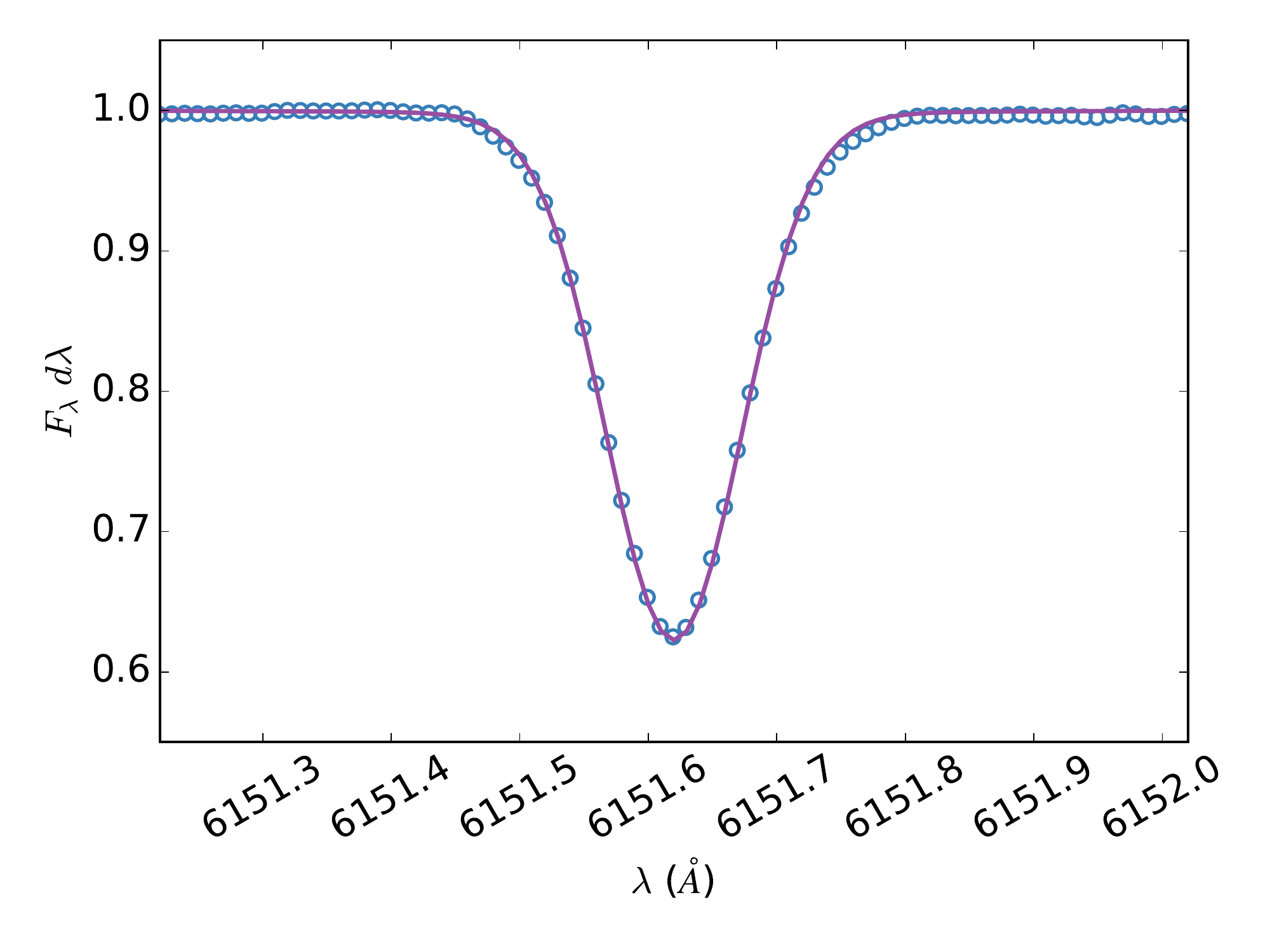}
\caption{Example of line profile fitting for the Fe I feature at 6151.62 \AA\ in the spectrum of the Sun. The continuous curve is the synthetic spectrum, and the open circles are the observed data.}
\label{sun_fit}
\end{figure}

We estimate $v \sin{i}$ for our sample of solar twins using the 2014 version of MOOG Synth \citep{1973PhDT.......180S}, adopting stellar atmosphere models by \citet{2004astro.ph..5087C}, with interpolations between models performed automatically by the Python package qoyllur-quipu\footnote{\footnotesize{Available at \url{https://github.com/astroChasqui/q2}}} \citep[see][]{2014A&A...572A..48R}. The instrumental broadening is taken into account by the spectral synthesis. We used the stellar parameters from \citet{2016A&A...590A..32T} and microturbulence velocities from \citet{2014A&A...572A..48R}. Macroturbulence velocities ($v_{\mathrm{macro}}$) were calculated by scaling the solar values, line by line (see Sect. \ref{vmacro_det}). Estimation of the rotational velocities was performed with our own algorithm\footnote{\footnotesize{Available at \url{https://github.com/RogueAstro/PoWeRS}}} that makes automatic measurements for all spectral lines for each star. We applied fine tuning corrections by eye for the non-satisfactory automatic line profile fittings, and quote $v \sin{i}$ as the mean of the values measured for the five lines. See Sects. \ref{vmacro_det} and \ref{vsini_det} for a detailed description on rotational velocities estimation as well as their uncertainties. Fig. \ref{sun_fit} shows an example of spectral line fitting for one feature in the Sun.

\subsection{Macroturbulence velocities}\label{vmacro_det}

We tested the possibility of measuring $v_{\mathrm{macro}}$ (radial-tangential profile) simultaneously with $v \sin{i}$, but even when using the extremely high-resolution spectra of HARPS, it is difficult to disentangle these two spectral line broadening processes, which is probably due to the low values of these velocities. Macroturbulence has a stronger effect on the wings of the spectral lines, but our selection of clean lines still has some contamination that requires this high-precision work to be done by eye. Some stars show more contamination than others, complicating the disentaglement. Fortunately, the variation of macroturbulence with effective temperature and luminosity is smooth \citep{2005oasp.book.....G}, so that precise values of $v_{\mathrm{macro}}$ could be obtained by a calibration. Thus we adopted a relation that fixes macroturbulence velocities in order to measure $v \sin{i}$ with high precision using an automatic code, which provides the additional benefits of reproducibility and lower subjectivity.

The macroturbulence velocity is known to vary for different spectral lines \citep{2005oasp.book.....G}, so for our high-precision analysis, we do not adopt a single value for each star. Instead, we measure the $v_{\mathrm{macro}}$ for the Sun in each of the spectral lines from Table \ref{lines}, and use these values to scale the $v_{\mathrm{macro}}$ for all stars in our sample using the following equation\footnote{\footnotesize{In the future, it should be possible to calibrate macroturbulence velocities using 3D hydrodynamical stellar atmosphere models \citep[e.g.,][]{2013A&A...557A..26M} by using predicted 3D line profiles (without rotational broadening) as observations and determine which value of $v_{\rm{macro}}$ is needed to reproduce them with 1D model atmospheres.}}:

\begin{eqnarray}
v_{\mathrm{macro},\lambda}^{*} = v_{\mathrm{macro},\lambda}^{\odot} - 0.00707\ T_{\mathrm{eff}} + 9.2422 \times 10^{-7}\ T_{\mathrm{eff}}^2 \nonumber \\ + 10.0 + k_1 \left(\log{g} - 4.44\right) + k_2 \\
\equiv f(T_{\mathrm{eff}}) + k_1\left(\log{g} - 4.44\right) + k_2 \nonumber
\label{vmacro_eq}
\end{eqnarray}where $v\rm_{macro,\lambda}^{\odot}$ is the macroturbulence velocity of the Sun for a given spectral line, $T_{\mathrm{eff}}$ and $\log{g}$ are, respectively, the effective temperature and gravity of a given star, $k_1$ is a proportionality factor for $\log{g}$ and $k_2$ is a small correction constant.

This formula is partly based on the relation derived by \citet{2012A&A...543A..29M} (Eq. E.1 in their paper) from the trend of macroturbulence with effective temperature in solar-type stars described by \citet{2005oasp.book.....G}. The $\log{g}$-dependent term (a proxy for luminosity) comes from the empirical relation derived by \citet{2014MNRAS.444.3592D} (Eq. 8 in their paper), and is based on spectroscopic measurements of $v_{\mathrm{macro}}$ of \textit{Kepler} stars, which were disentangled from $v \sin{i}$ using asteroseismic estimates of the projected rotational velocities. Doyle et al. obtained a value for the proportionality factor $k_1$ of -$2.0$. However their uncertainties on $v_{\mathrm{macro}}$ were of the order of 1.0 km s$^{-1}$. Thus, we decided to derive our own values of $k_1$ and $k_2$ by simultaneously measuring $v_{\mathrm{macro}}$ and $v \sin{i}$ of a sub-sample of solar twins.

\begin{figure}[!t]
   \centering
   \includegraphics[width=\hsize]{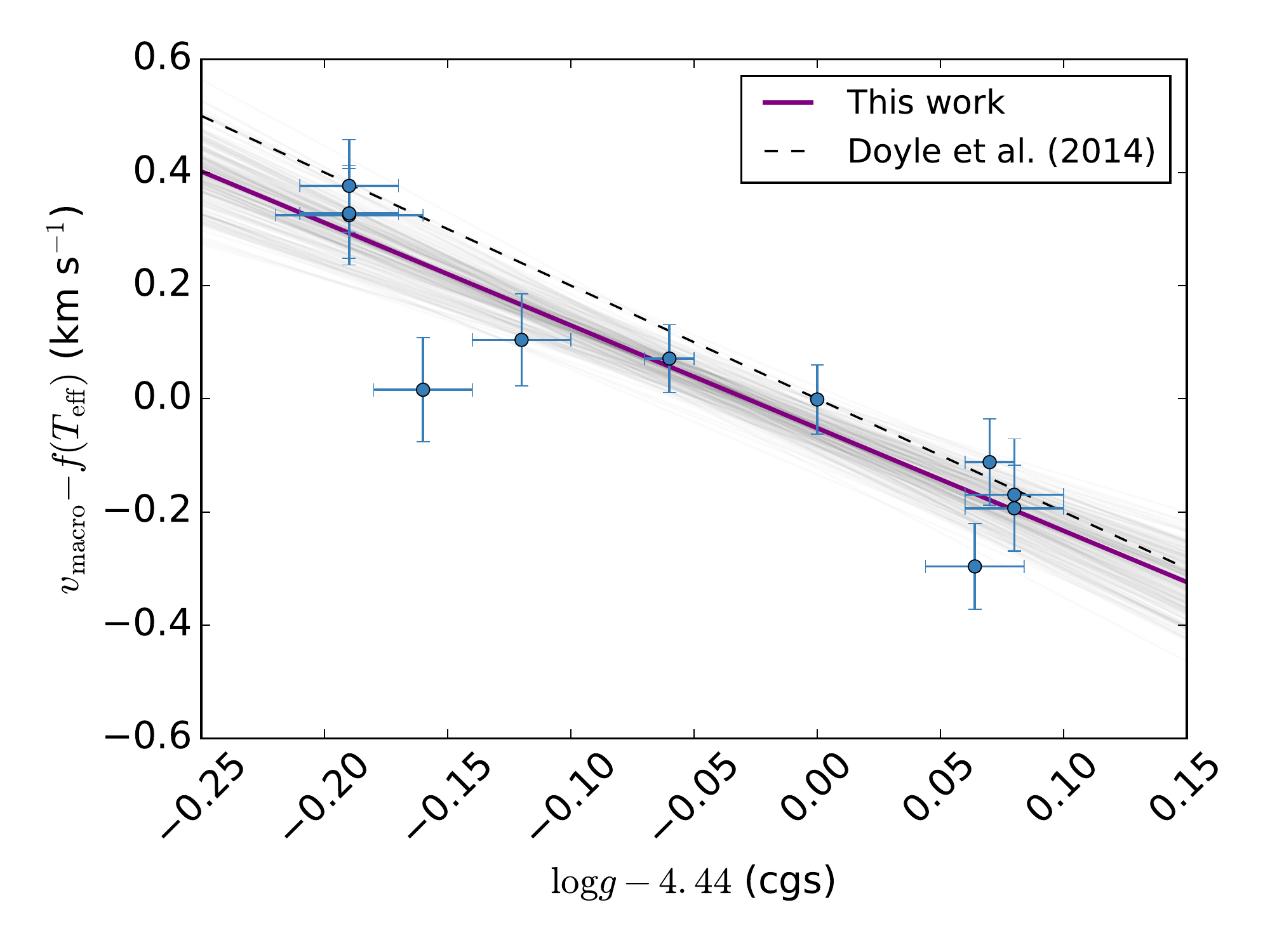}
   \caption{Linear relation between $v_{\mathrm{macro}}$ and $\log{g}$ (a proxy for luminosity) for the stars on Table \ref{vmacro_stars}. See the definition of $f(T_{\mathrm{eff}})$ in Sect. \ref{vmacro_det}. The orange continuous line represents our determination of a proportionality coefficient of -1.81 and a vertical shift of -0.05 km s$^{-1}$. The black dashed line is the coefficient found by \citet{2014MNRAS.444.3592D}. The light grey region is a composition of 200 curves with parameters drawn from a multivariate gaussian distribution. The Sun is located at the origin.}
   \label{vmacro_logg}
   \end{figure}

This sub-sample was chosen to contain only single stars or visual binaries mostly in the extremes of $\log{g}$ ($4.25$ -- $4.52$) in our entire sample. We assume these values to have a linear relationship with $v_{\mathrm{macro}}$ inside this short interval of $\log{g}$. We used as a first guess the values of $v \sin{i}$ and $v_{\mathrm{macro}}$ from a previous, cruder estimation we made, and performed line profile fits by eye using MOOG Synth. The velocities in Table \ref{vmacro_stars} are the median of the values measured for each line and their standard error. Note that these $v \sin{i}$ are not consistently measured in the same way that the final results are. The rotational velocity broadening was calculated by our own code (see Sect. \ref{vsini_det} for details). By performing a linear fit in the $v_{\mathrm{macro}} - f(T_{\mathrm{eff}})$ vs. $\log{g}-4.44$ relation ($f$ comprises all the $T_{\mathrm{eff}}$-dependent terms, the macroturbulence velocity of the Sun and the known constant on Eq. \ref{vmacro_eq}), we obtain that $k_1 = -1.81 \pm 0.26$ and $k_2 = -0.05 \pm 0.03$ (see Fig. \ref{vmacro_logg}). For the stars farthest from the Sun in $\log{g}$ from our sample, these values of $k_1$ and $k_2$ would amount to differences of up to $\pm 0.4$ km s$^{-1}$ in their macroturbulence velocities, therefore it is essential to consider the luminosity effect on $v\rm_{macro}$ for accurate $v \sin{i}$ determinations.

\begin{table}[h]
\begin{center}
\caption{Simultaneous measurements of rotational and macroturbulence velocities of stars in the extremes of $\log{g}$ from our sample of solar twins.}
\begin{tabular}{lcccc}
\hline \hline\\[-2ex]
Star & $v \sin{i}$ & $v_{\mathrm{macro}}$ & $T\rm_{eff}$ & $\log{g}$ \\
 & (km s$^{-1}$) & (km s$^{-1}$) & & \\ \hline
HIP 115577 & $0.95 \pm 0.05$ & $3.35 \pm 0.09$ & 5699 & 4.25 \\
HIP 65708 & $1.20 \pm 0.09$ & $3.55 \pm 0.08$ & 5755 & 4.25 \\
HIP 74432 & $1.40 \pm 0.03$ & $3.35 \pm 0.08$ & 5684 & 4.25 \\
HIP 118115 & $1.40 \pm 0.10$ & $3.43 \pm 0.09$ & 5808 & 4.28 \\
HIP 68468 & $1.75 \pm 0.07$ & $3.70 \pm 0.08$ & 5857 & 4.32 \\
HIP 41317 & $1.55 \pm 0.03$ & $3.10 \pm 0.06$ & 5700 & 4.38 \\
Sun & $1.75 \pm 0.07$ & $3.30 \pm 0.06$ & 5777 & 4.44 \\
HIP 105184 & $2.50 \pm 0.03$ & $3.21 \pm 0.08$ & 5833 & 4.50 \\
HIP 10175 & $1.55 \pm 0.06$ & $3.05 \pm 0.08$ & 5738 & 4.51 \\
HIP 114615 & $2.20 \pm 0.03$ & $3.25 \pm 0.08$ & 5816 & 4.52 \\
HIP 3203 & $3.90 \pm 0.03$ & $3.40 \pm 0.10$ & 5850 & 4.52 \\
\hline
\end{tabular}
\label{vmacro_stars}
\end{center}
\end{table}

To obtain the macroturbulence velocities for the Sun to use in Eq. \ref{vmacro_eq}, we forced the rotational velocity of the Sun to 1.9 km s$^{-1}$ \citep{1970SoPh...12...23H}, and then estimated values of $v\rm_{macro, \lambda}^{\odot}$ by fitting each line profile using MOOG Synth, and the results are shown in Table \ref{lines}. We estimate the error in determining $v\rm_{macro,\lambda}^{\odot}$ to be $\pm 0.1$ km s$^{-1}$. Since Eq. \ref{vmacro_eq} is an additive scaling, the error for $v\rm_{macro}$ of all stars is the same as in the Sun\footnote{\footnotesize{The uncertainties in stellar parameters have contributions that are negligible compared to the ones introduced by the error in $v_{\mathrm{macro}}$.}}.

\subsection{Rotational velocities}\label{vsini_det}

Our code takes as input the list of stars and their parameters (effective temperature, surface gravity, metallicity and microturbulence velocities obtained on Paper I), their spectra and the spectral line list in MOOG-readable format. For each line in a given star, the code automatically corrects the spectral line shift and the continuum. The first is done by fitting a second order polynomial to the kernel of a line and estimating what distance the observed line center is from the laboratory value. Usually, the spectral line shift corrections were of the order of $10^{-2}$ \AA, corresponding to 0.5 km s$^{-1}$ in the wavelength range we worked on. This is a reasonable shift that likely arises from a combination of granulation and gravitational redshift effects, which are of similar magnitude. The continuum correction for each line is defined as the value of a multiplicative factor that sets the highest flux inside a radius of 2.5 \AA\ around the line center to 1.0. The multiplicative factor usually has a value inside the range $1.000 \pm 0.002$.

The code starts with a range of $v \sin{i}$ and abundances and optimizes these two parameters through a series of iterations that measure the least squares difference between the observed line and the synthetic line (generated with MOOG synth). Convergence is achieved when the difference between the best solution and the previous one, for both $v \sin{i}$ and abundance, is less than 1\%. Additionally, the code also forces at least 10 iterations in order to avoid falling into local minima.

One of the main limitations of MOOG Synth for our analysis is that it has a "quantized" behavior for $v \sin{i}$: the changes in the synthetic spectra occur most strongly in steps of 0.5 km s$^{-1}$. This behavior is not observed in varying the macroturbulence velocities. Therefore, we had to incorporate a rotational broadening routine in our code that was separated from MOOG. We used the Eq. 18.14 from \citet{2005oasp.book.....G}, in velocity space, to compute the rotational profile\footnote{\footnotesize{This is the same recipe adopted by the radiative transfer code MOOG.}}:

\begin{equation}
G(v) = \frac{2(1-\epsilon)\left[ 1-(v/v\rm_L)^2 \right]^{1/2} + \frac{1}{2} \pi \epsilon \left[ 1-(v/v\rm_L)^2 \right]}{\pi v\rm_L (1-\epsilon/3)}\mathrm{,}
\end{equation}where $v\rm_L$ is the projected rotational velocity and $\epsilon$ is the limb darkening coefficient (for which we adopt the value 0.6). The rotational profile $G(v)$ is then convolved with MOOG's synthetic profiles (which were generated with $v \sin{i}$ = 0).

The total uncertainties in rotational velocities are obtained from the quadratic sum of the standard error of the five measurements and an uncertainty of 0.1 km s$^{-1}$ introduced by the error in macroturbulence velocities. Systematic errors in the calculation of $v\rm_{macro,\lambda}$ for the stars do not significantly contribute to the $v \sin{i}$ uncertainties.

\begin{figure}[!t]
\centering
\includegraphics[width=\hsize]{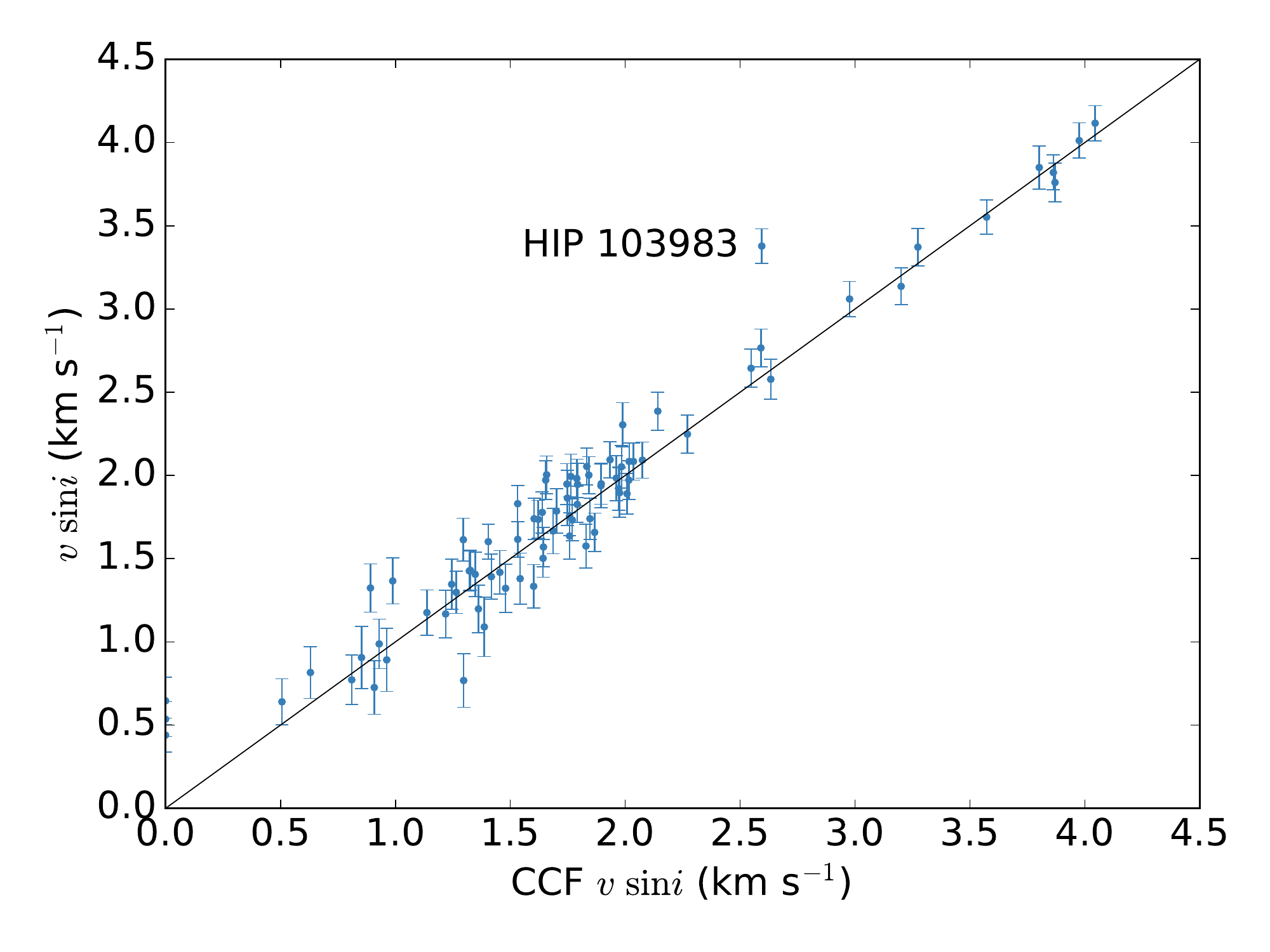}
\caption{Comparison between our estimated values of $v \sin{i}$ (y-axis) and the ones inferred from the cross-correlation funcion FWHM (x-axis). The spread around the 1:1 relation (black line) is $\sigma = 0.20$ km s$^{-1}$.}
\label{ccf}
\end{figure}

Some of the stars in the sample show very low rotational velocities, most probably due to the effect of projection (see left panel of Fig. \ref{sku}). The achieved precision is validated by comparison with the values of the full-width at half maximum (FWHM) measured by the cross-correlation function (CCF) from the data reduction pipeline, with the effects of macroturbulence subtracted (see Fig. \ref{ccf}). The spectroscopic binary star HIP 103983 has an unusually high $v \sin{i}$ when compared to the CCF FWHM, and a verification of its spectral line profiles reveals the presence of distortions that are the most probably caused by mis-measurement of rotational velocity (contamination of the combined spectrum by a companion -- observations range from October 2011 to August 2012). We obtained a curve fit for the $v \sin{i}$ vs. CFF FWHM (km s$^{-1}$) using a similar relation as used by \citet{2001A&A...375..851M, 2004A&A...426.1021P, 2007A&A...475.1003H}, which resulted in the following calibration: $v \sin{i} = \sqrt{(0.73 \pm 0.02) \left[\mathrm{FWHM}^2 - v_{\mathrm{macro}}^2 - (5.97 \pm 0.01)^2\right]}$ km s$^{-1}$ \citep[estimation performed with the MCMC code \texttt{emcee}\footnote{\footnotesize{Available at \url{http://dan.iel.fm/emcee/current/}}}][]{2013PASP..125..306F}. The scatter between the measured $v \sin{i}$ and the ones estimated from CCF is $\sigma = 0.20$ km s$^{-1}$ (excluding the outlier HIP 103983). The typical uncertainty in the rotational velocities we obtain with our method -- line profile fitting with extreme high resolution spectra -- is 0.12 km s$^{-1}$, which implies that the average error of the CCF FWHM $v \sin{i}$ scaling is 0.16 km s$^{-1}$, which could be significantly higher if the broadening by $v_\mathrm{macro}$ is not accounted for.

\section{Binary stars}

We identified 16 spectroscopic binaries (SB) in our sample of 82 solar twins by analyzing their radial velocities; some of these stars are reported as binaries by \citet{2014AJ....147...86T, 2014AJ....147...87T, 2001AJ....121.3224M, 2015ApJ...802...37B}. We did not find previous reports of multiplicity for the stars HIP 30037, HIP 62039 and HIP 64673 in the literature. Our analysis of variation in the HARPS radial velocities suggest that the first two are probable SBs, while the latter is a candidate. No binary shows a double-lined spectrum, but HIP 103983 has distortions that could be from contamination by a companion. The star HIP 64150 is a Sirius-like system with a directly observed white dwarf companion \citep{2013ApJ...774....1C,2014ApJ...783L..25M}. The sample from Paper I contains another SB, HIP 109110, for which we could not reliably determine the $v \sin{i}$ due to strong contamination in the spectra, possibly caused by a relatively bright companion. Thus, we did not include this star in our sample.

\begin{figure*}[!ht]
\centering
\begin{tabular}{cc}
\includegraphics[width=0.48\textwidth]{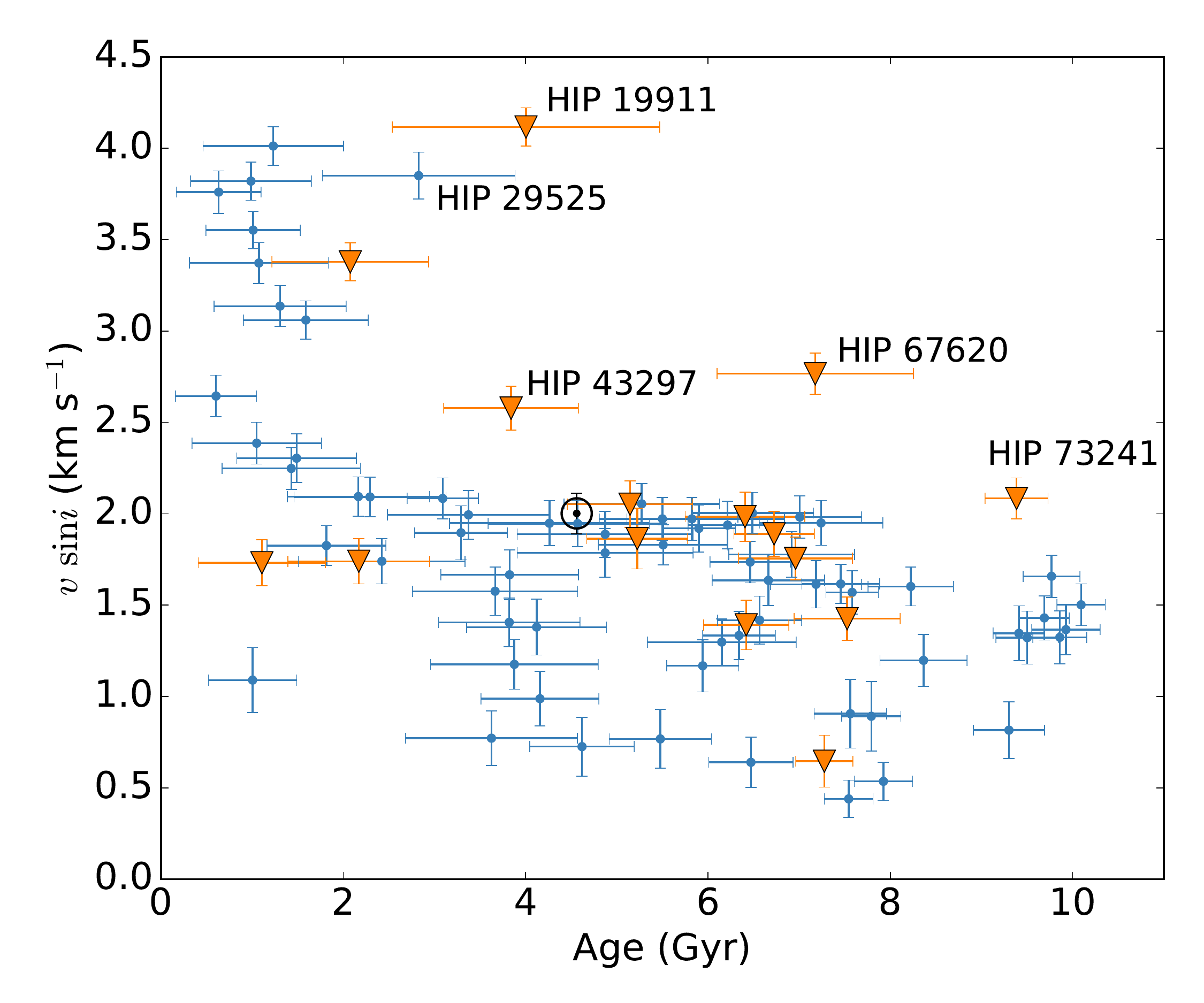} & \includegraphics[width=0.48\textwidth]{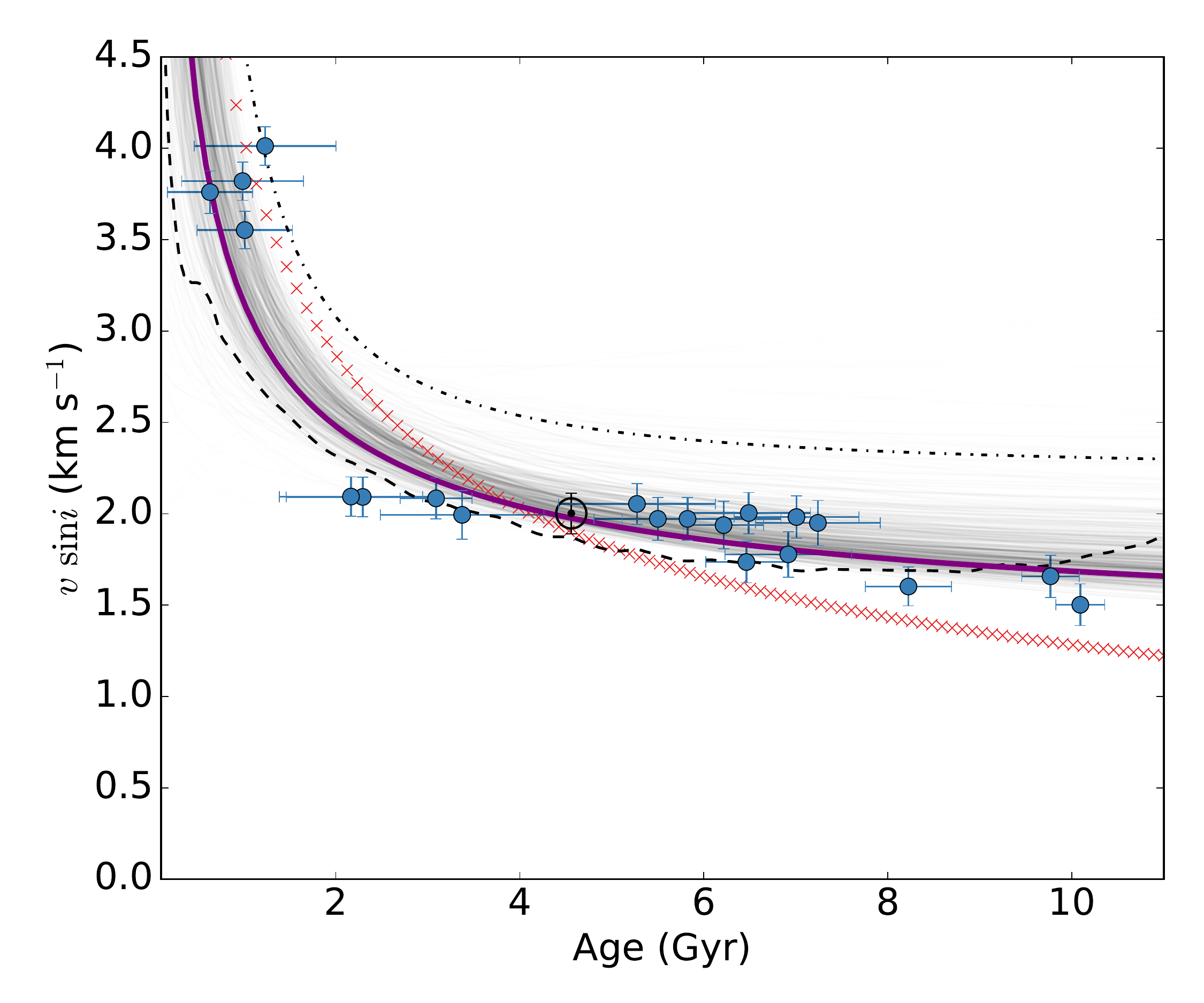} \\
\end{tabular}
\caption{Projected rotational velocity of solar twins in function of their age. The Sun is represented by the symbol $\odot$. Left panel: all stars of our sample; the orange triangles are spectroscopic binaries, blue circles are the \textit{selected sample} and the blue dots are the remaining non-spectroscopic binaries. Right panel: the rotational braking law; the purple continous curve is our relation inferred from fitting the \textit{selected sample} (blue circles) of solar twins with the form $v \sin{i} = v_\mathrm{f} + m\ t^{-b}$, where $t$ is the stellar age, and the fit parameters are $v_\mathrm{f} = 1.224 \pm 0.447$, $m = 1.932 \pm 0.431$ and $b = 0.622 \pm 0.354$, with $v_\mathrm{f}$ and $b$ highly and positively correlated. The light grey region is composed of 300 curves that are created with parameters drawn from a multivariate gaussian distribution defined by the mean values of the fit parameters and their covariance matrix. Skumanich's law (red $\times$ symbols, calibrated for $v^\odot_{\mathrm{rot}} = 1.9$ km s$^{-1}$) and the rotational braking curves proposed by \citet[][black dashed curve, smoothed]{2014ApJ...790L..23D} and \citet[][black dot-dashed curve]{2004A&A...426.1021P} are plotted for comparison.}
\label{sku}
\end{figure*}

Of these 16 spectroscopic binaries, at least four of them (HIP 19911, 43297, 67620 and 73241) show unusually high $v \sin{i}$ (see the left panel of Fig. \ref{sku}). These stars also present other anormalities, such as their [Y/Mg] abundances \citep{2016A&A...590A..32T} and magnetic activity \citep{2014A&A...572A..48R, F16inprep}. The solar twin blue straggler HIP 10725 \citep{2015A&A...584A.116S}, which is not included in our sample, also shows a high $v \sin{i}$ for its age. We find that five of the binaries have rotational velocities below the expected for Sun-like stars, but this is most likely an effect of projection of the stars' rotational axes. For the remaining binaries, which follow the rotational braking law, it is again difficult to disentangle this behavior from the $\sin{i}$, and a statistical analysis is precluded by the low numbers involved. Tidal interactions between companions that could potentially enhance rotation depend on binary separation, which is unknown for most of these stars. They should be regular rotators, since they do not show anormalities in chromospheric activity \citep{F16inprep} or [Y/Mg] abundances \citep{2016A&A...590A..32T}.

Based on the information that at least 25\% of the spectroscopic binaries in our sample show higher rotational velocities than expected for single stars, we conclude that stellar multiplicity is an important enhancer of rotation in Sun-like stars. Blue stragglers are expected to have a strong enhancement on rotation due to injection of angular momentum from the donor companion.


\section{The rotational braking law}\label{br_law}

In order to correctly constrain the rotational braking, we removed from this analysis all the spectroscopic binaries. The non-SB HIP 29525 displays a $v \sin{i}$ much higher than expected ($3.85 \pm 0.13$ km s$^{-1}$), but it is likely that this is due to an overestimated isochronal age ($2.83 \pm 1.06$ Gyr). Because it is a clear outlier in our results, we decided to not include HIP 29525 in the rotational braking determination. \citet{2010A&A...521A..12M} found X-ray and chromospheric ages of 0.55 and 0.17 Gyr, respectively, for HIP 29525. We then divided the remaining 65 stars and the Sun in bins of 2 Gyr, and removed from this sample all the stars which were below the 70th percentile of $v \sin{i}$ in each bin\footnote{\footnotesize{By doing a simple simulation with angles $i$ drawn from a flat distribution between 0 and $\pi/2$, we verify that 30\% of the stars should have $\sin{i}$ above 0.9.}}. This allowed us to select the stars that had the highest chance of having $\sin{i}$ above $0.9$. In total, 21 solar twins and the Sun compose what we hereafter reference as the \textit{selected sample}. Albeit this sub-sample is smaller, it has the advantage of mostly removing uncertainties on the inclination angle of the stellar rotation axes\footnote{\footnotesize{This procedure can also allow for unusually fast-rotating stars (although rare) with $\sin {i}$ below 0.9 to leak into our sample.}}. We stress that the only reason we can select the most probable edge-on rotating stars ($i = \pi/2$) is because we have a large sample of solar twins in the first place.

We then proceeded to fit a general curve to the \textit{selected sample} (see Fig. \ref{sku}) using the method of orthogonal distance regression \citep[ODR,][]{boggs1990orthogonal}, which takes into account the uncertainties on both $v \sin{i}$ and ages. This curve is a power law plus constant of the form $v = v_\mathrm{f} + m\ t^{-b}$ \citep[the same chromospheric activity and $v \sin{i}$ vs. age relation used by][]{2004A&A...426.1021P, 2009IAUS..258..395G}, with $v$ (rotational velocity) and $v\rm_f$ (asymptotic velocity) in km s$^{-1}$ and $t$ (age) in Gyr.

We find that the best fit parameters are $v_\mathrm{f} = 1.224 \pm 0.447$, $m = 1.932 \pm 0.431$ and $b = 0.622 \pm 0.354$ (see right panel of Fig. \ref{sku}). These large uncertainties are likely due to: i) the strong correlation between $v_\mathrm{f}$ and $b$; and ii) the relatively limited number of datapoints between 1 and 4 Gyr, where the parameter is most effective in changing the values of $v$. This limitation is also present in past studies \citep[e.g.,][]{2016Natur.529..181V, 2003ApJ...586..464B, 2004A&A...426.1021P, 2008ApJ...687.1264M, 2014A&A...572A..34G, 2016A&A...587A.105A}. On the other hand, our sample is the largest comprising solar twins, and therefore should produce more reliable results. With more datapoints, we could be able to use 1 Gyr bins instead of 2 Gyr in order to select the fastest rotating stars, which would result in a better sub-sample for constraining the rotational evolution for young stars.

The relation we obtain is in contrast with some previous studies on modelling the rotational braking \citep{2001ApJ...561.1095B, 2003ApJ...586..464B, 2015A&A...584A..30L} which either found or assumed that the Skumanich's law explains well the rotational braking of Sun-like stars. The conclusions by \citet{2016Natur.529..181V} limit the range of validation up to approximately the solar age (4 Gyr) for stars with solar mass. When we enforce the Skumanich's power law index $b = 1/2$, we obtain a worse fit between the ages 2 and 4 Gyr (and, not surprisingly, also after the solar age).

Our data and the rotational braking law that results from them show that the Sun is a normal star regarding its rotational velocity when compared to solar twins. However, they do not agree with a regular Skumanich's law \citep[][red $\times$ symbols in Fig. \ref{sku}]{2007ApJ...669.1167B}. We find a better agreement with the model proposed by \citet[][black dashed curve in Fig. \ref{sku}]{2014ApJ...790L..23D}, especially for stars older than 2 Gyr. This model is thoroughly described in Appendix A of \citet{2012A&A...548L...1D}. In summary, it uses an updated treatment of the instabilities relevant to the transport of angular momentum according to \citet{1992A&A...265..115Z} and \citet{1997A&A...317..749T}, with an initial angular momentum for the Sun $J_0 = 1.63 \times 10^{50}$ g cm$^2$ s$^{-1}$. Its corresponding rotational braking curve is computed using the output radii of the model, which vary from $\sim$1 R$_\odot$ at the current solar age to 1.57 R$_\odot$ at the age of 11 Gyr, and it changes significantly if we use a constant radius $R = 1$ R$_\odot$, resulting in a more Skumanich-like rotational braking.

Our result agrees with the chromospheric activity vs. age behavior for solar twins obtained by \citet{2014A&A...572A..48R}, in which a steep decay of the $R'\rm_{HK}$ index during the first 4 Gyr was deduced (see Fig. 11 in their paper). The study by \citet{2004A&A...426.1021P} also suggests a steeper power-law index ($b = 1.47$) than Skumanich's ($b_\mathrm{S} = 1/2$) in the rotational braking law derived from young open clusters, the Sun and M 67. However, as seen in Fig. \ref{sku}, their relation significantly overestimates the rotational velocities of stars, especially for those older than 2 Gyr. This is most probably caused by other line broadening processes, mainly the macroturbulence, which were not considered in that study. As we saw in Sect. \ref{vmacro_det}, those introduce important effects that are sometimes larger than the rotational broadening. Moreover, a CCF-only analysis tends to produce more spread in the $v \sin{i}$ than the more detailed analysis we used.

The rotational braking law we obtain produces a similar outcome to that achieved by \citet{2016Natur.529..181V} for stars older than the Sun (a weaker rotational braking law after solar age than previously suggested). Our data also requires a different power law index than Skumanich's index for stars younger than the Sun, one that accounts for an earlier decay of rotational velocities up to 2 Gyr.

The main sequence spin-down model by \citet{1988ApJ...333..236K} states that, for constant moment of inertia and radius during the main sequence, we would have

\begin{equation}
 v_{\mathrm{eq}} \propto t^{-3/(4an)} \mathrm{,}
 \label{kawaler}
\end{equation}where $v_{\mathrm{eq}}$ is the rotational velocity at the equator and $a$ and $n$ are parameters that measure the dependence on rotation rate and radius, respectively (see Eqs. 7, 8 and 12 in their paper). If we assume a dipole geometry for the stellar magnetic field ($B_\mathrm{r} \propto B_0 r^3$), then $n = 3/7$. Furthermore, assuming that $a = 1$, then Eq. \ref{kawaler} results in $v_{\mathrm{eq}} \propto t^{-7/4} = t^{-1.75}$. Skumanich's law ($v_{\mathrm{eq}} \propto t^{-0.5}$) is recovered for $n = 3/2$, which is close to the case of a purely radial field ($n = 2$, $v_{\mathrm{eq}} \propto t^{-0.38}$). A more extensive exploration of the configuration and evolution of magnetic fields of solar twins is outside the scope of this paper, but our results suggest that the rotational rotational braking we observe on this sample of solar twins stems from a magnetic field with an intermediate geometry between dipole and purely radial.


\section{Conclusions}

We analyzed the rotational velocities of 82 bright solar twins in the Southern Hemisphere and the Sun using extremely high resolution spectra. Radial velocities revealed that our sample contained 16 spectroscopic binaries, three of which (HIP 30037, 62039, 64673) were not listed as so in the literature. At least five of these stars show an enhancement on their measured $v \sin{i}$, which is probably caused by interaction with their close-by companions. They also present other anomalies in chemical abundances and chromospheric activities. We did not clearly identify non-spectroscopic binary stars with unusually high rotational velocities for their age.

In order to better constrain the rotational evolution of the solar twins, we selected a subsample of stars with higher chances of having their rotational axis inclination close to $\pi/2$ (almost edge-on). We opted to use carefully measured isochronal ages for these stars because it is the most reliable method available for this sample. We finally conclude that the Sun seems to be a common rotator, within our uncertainties, when compared to solar twins, therefore it can be used to calibrate stellar models.

Moreover, we have found that Skumanich's law does not describe well the rotation evolution for solar twins observed in our data, a discrepancy that is stronger after the solar age. Therefore, we propose a new rotational braking law that supports the weakened braking after the age of the Sun, and comes with a earlier decay in rotational velocities up to 2 Gyr than the classical Skumanich's law. Interestingly, it also reveals an evolution that is more similar to the magnetic activity evolution observed in Sun-like stars, which sees a steep decay in the first 3 Gyr and flattens near the solar age. Additionally, we suggest that more high-precision spectroscopic observations of solar twins younger and much older than the Sun could help us better constrain the rotational evolution of solar-like stars.
\begin{acknowledgements}
      LdS thanks CAPES and FAPESP, grants no. 2014/26908-1 and 2016/01684-9 for support. JM thanks for support by FAPESP (2012/24392-2). LS acknowledges support by FAPESP (2014/15706-9). We also would like to thank the anonymous referee for the valuable comments that significantly improved this manuscript.
\end{acknowledgements}

\bibliographystyle{aa}
\bibliography{rot_ST}

\Online

\newgeometry{left=2.5cm, bottom=2cm}
\small

\begin{center}
\begin{longtable}{l|cc|cc|cc|cc|cc|cc|c}
\caption{Ages, the measured $v \sin{i}$ and stellar parameters of the 82 solar twins and the Sun.}\\
\hline \hline \\[-2ex]
\multicolumn{1}{c|}{Star} &
\multicolumn{1}{c}{Age} &
\multicolumn{1}{c|}{$\sigma$} &
\multicolumn{1}{c}{$v \sin{i}$} &
\multicolumn{1}{c|}{$\sigma$} &
\multicolumn{1}{c}{[Fe/H]} &
\multicolumn{1}{c|}{$\sigma$} &
\multicolumn{1}{c}{$T\rm_{eff}$} &
\multicolumn{1}{c|}{$\sigma$} &
\multicolumn{1}{c}{log $g$} &
\multicolumn{1}{c|}{$\sigma$} &
\multicolumn{1}{c}{$v\rm_{t}$} &
\multicolumn{1}{c|}{$\sigma$} &
\multicolumn{1}{c}{$v\rm_{macro}$} \\
\multicolumn{1}{c|}{} &
\multicolumn{2}{c|}{(Gyr)} &
\multicolumn{2}{c|}{(km s$^{-1}$)} &
\multicolumn{2}{c|}{(dex)} &
\multicolumn{2}{c|}{(K)} &
\multicolumn{2}{c|}{(cgs)} &
\multicolumn{2}{c|}{(km s$^{-1}$)} &
\multicolumn{1}{c}{(km s$^{-1}$)} \\

\hline
\endfirsthead

\multicolumn{11}{c}{\footnotesize{{\slshape{{\tablename} \thetable{}}} - Continuation}}\\

\hline \hline\\[-2ex]

\multicolumn{1}{c|}{Star} &
\multicolumn{1}{c}{Age} &
\multicolumn{1}{c|}{$\sigma$} &
\multicolumn{1}{c}{$v \sin{i}$} &
\multicolumn{1}{c|}{$\sigma$} &
\multicolumn{1}{c}{[Fe/H]} &
\multicolumn{1}{c|}{$\sigma$} &
\multicolumn{1}{c}{$T\rm_{eff}$} &
\multicolumn{1}{c|}{$\sigma$} &
\multicolumn{1}{c}{log $g$} &
\multicolumn{1}{c|}{$\sigma$} &
\multicolumn{1}{c}{$v\rm_{t}$} &
\multicolumn{1}{c|}{$\sigma$} &
\multicolumn{1}{c}{$v\rm_{macro}$} \\
\multicolumn{1}{c|}{} &
\multicolumn{2}{c|}{(Gyr)} &
\multicolumn{2}{c|}{(km s$^{-1}$)} &
\multicolumn{2}{c|}{(dex)} &
\multicolumn{2}{c|}{(K)} &
\multicolumn{2}{c|}{(cgs)} &
\multicolumn{2}{c|}{(km s$^{-1}$)} &
\multicolumn{1}{c}{(km s$^{-1}$)} \\

\hline
\endhead

\hline \multicolumn{11}{l}{{Continued on next page}} \\
\endfoot
\hline

\endlastfoot

HIP 1954 & 4.87 & 0.97 & 1.79 & 0.13 & -0.068 & 0.006 & 5717 & 5 & 4.46 & 0.02 & 0.96 & 0.02 & 2.90  \\
HIP 3203$^\dagger$ & 0.99 & 0.66 & 3.82 & 0.11 & -0.087 & 0.008 & 5850 & 10 & 4.52 & 0.02 & 1.16 & 0.02 & 3.27  \\
HIP 4909$^\dagger$ & 1.23 & 0.77 & 4.01 & 0.11 & 0.028 & 0.008 & 5854 & 10 & 4.50 & 0.02 & 1.12 & 0.02 & 3.33  \\
HIP 5301$^\dagger$ & 6.49 & 0.67 & 2.00 & 0.12 & -0.064 & 0.004 & 5728 & 5 & 4.42 & 0.02 & 0.97 & 0.01 & 3.01  \\
HIP 6407* & 1.49 & 0.66 & 2.30 & 0.13 & -0.068 & 0.007 & 5764 & 8 & 4.52 & 0.01 & 0.97 & 0.02 & 2.96  \\
HIP 7585 & 3.29 & 0.51 & 1.90 & 0.15 & 0.095 & 0.005 & 5831 & 5 & 4.43 & 0.01 & 1.02 & 0.01 & 3.37  \\
HIP 8507 & 3.63 & 0.94 & 0.77 & 0.15 & -0.096 & 0.006 & 5725 & 6 & 4.49 & 0.02 & 0.99 & 0.02 & 2.88  \\
HIP 9349 & 1.43 & 0.76 & 2.25 & 0.11 & 0.009 & 0.007 & 5810 & 8 & 4.50 & 0.02 & 1.07 & 0.02 & 3.16  \\
HIP 10175 & 1.82 & 0.65 & 1.83 & 0.11 & -0.007 & 0.005 & 5738 & 7 & 4.51 & 0.01 & 0.96 & 0.01 & 2.89  \\
HIP 10303 & 5.48 & 0.56 & 0.77 & 0.16 & 0.106 & 0.004 & 5725 & 4 & 4.40 & 0.01 & 0.98 & 0.01 & 3.04  \\
HIP 11915 & 4.16 & 0.65 & 0.99 & 0.15 & -0.059 & 0.004 & 5760 & 4 & 4.46 & 0.01 & 0.97 & 0.01 & 3.05  \\
HIP 14501 & 9.93 & 0.37 & 1.37 & 0.14 & -0.133 & 0.005 & 5728 & 7 & 4.29 & 0.02 & 1.03 & 0.01 & 3.25  \\
HIP 14614$^\dagger$ & 5.82 & 1.02 & 1.97 & 0.12 & -0.099 & 0.008 & 5784 & 9 & 4.42 & 0.03 & 1.03 & 0.02 & 3.21  \\
HIP 15527 & 7.92 & 0.32 & 0.54 & 0.11 & -0.051 & 0.005 & 5785 & 5 & 4.32 & 0.01 & 1.05 & 0.01 & 3.40  \\
HIP 18844 & 7.46 & 0.43 & 1.62 & 0.11 & 0.016 & 0.004 & 5736 & 5 & 4.36 & 0.02 & 0.99 & 0.01 & 3.15  \\
HIP 19911* & 4.00 & 1.47 & 4.12 & 0.11 & -0.070 & 0.011 & 5764 & 12 & 4.47 & 0.04 & 1.02 & 0.03 & 3.05  \\
HIP 22263 & 1.07 & 0.76 & 3.37 & 0.11 & 0.030 & 0.007 & 5840 & 8 & 4.50 & 0.02 & 1.08 & 0.02 & 3.27  \\
HIP 25670 & 4.12 & 0.77 & 1.38 & 0.15 & 0.057 & 0.005 & 5771 & 5 & 4.44 & 0.02 & 1.00 & 0.01 & 3.13  \\
HIP 28066 & 9.86 & 0.30 & 1.32 & 0.15 & -0.128 & 0.004 & 5733 & 5 & 4.29 & 0.01 & 1.05 & 0.01 & 3.27  \\
HIP 29432 & 5.51 & 0.71 & 1.83 & 0.11 & -0.096 & 0.005 & 5758 & 5 & 4.44 & 0.01 & 1.01 & 0.01 & 3.08  \\
HIP 29525 & 2.83 & 1.06 & 3.85 & 0.13 & -0.022 & 0.007 & 5737 & 7 & 4.49 & 0.02 & 1.12 & 0.02 & 2.92  \\
HIP 30037* & 6.96 & 0.62 & 1.76 & 0.12 & -0.011 & 0.004 & 5668 & 5 & 4.42 & 0.01 & 0.94 & 0.01 & 2.81  \\
HIP 30158 & 4.57 & 0.98 & 1.95 & 0.13 & 0.003 & 0.006 & 5702 & 5 & 4.46 & 0.02 & 0.94 & 0.02 & 2.85  \\
HIP 30476 & 9.69 & 0.27 & 1.43 & 0.12 & -0.022 & 0.004 & 5710 & 5 & 4.26 & 0.01 & 1.03 & 0.01 & 3.24  \\
HIP 30502$^\dagger$ & 7.01 & 0.68 & 1.98 & 0.12 & -0.076 & 0.006 & 5721 & 6 & 4.41 & 0.02 & 0.98 & 0.02 & 3.01  \\
HIP 33094$^\dagger$ & 10.09 & 0.27 & 1.50 & 0.12 & 0.043 & 0.005 & 5662 & 7 & 4.16 & 0.02 & 1.13 & 0.01 & 3.26  \\
HIP 34511$^\dagger$ & 3.37 & 0.89 & 1.99 & 0.13 & -0.103 & 0.006 & 5819 & 6 & 4.47 & 0.02 & 1.03 & 0.02 & 3.25  \\
HIP 36512 & 7.19 & 0.50 & 1.61 & 0.13 & -0.117 & 0.004 & 5737 & 4 & 4.41 & 0.01 & 0.99 & 0.01 & 3.06  \\
HIP 36515$^\dagger$ & 0.63 & 0.46 & 3.76 & 0.12 & -0.021 & 0.009 & 5847 & 12 & 4.54 & 0.02 & 1.17 & 0.02 & 3.23  \\
HIP 38072 & 1.31 & 0.72 & 3.14 & 0.11 & 0.058 & 0.007 & 5849 & 8 & 4.49 & 0.02 & 1.14 & 0.02 & 3.33  \\
HIP 40133$^\dagger$ & 5.50 & 0.39 & 1.97 & 0.12 & 0.128 & 0.004 & 5755 & 4 & 4.37 & 0.01 & 1.01 & 0.01 & 3.20  \\
HIP 41317$^\dagger$ & 8.22 & 0.47 & 1.60 & 0.11 & -0.068 & 0.004 & 5700 & 5 & 4.38 & 0.01 & 0.96 & 0.01 & 2.99  \\
HIP 42333$^\dagger$ & 1.01 & 0.52 & 3.55 & 0.10 & 0.138 & 0.008 & 5848 & 8 & 4.50 & 0.02 & 1.16 & 0.02 & 3.30  \\
HIP 43297* & 3.84 & 0.74 & 2.58 & 0.12 & 0.083 & 0.006 & 5702 & 5 & 4.46 & 0.01 & 0.99 & 0.02 & 2.85  \\
HIP 44713 & 7.58 & 0.29 & 1.57 & 0.12 & 0.088 & 0.005 & 5768 & 6 & 4.28 & 0.01 & 1.06 & 0.01 & 3.41  \\
HIP 44935$^\dagger$ & 6.22 & 0.43 & 1.94 & 0.13 & 0.058 & 0.005 & 5782 & 5 & 4.37 & 0.01 & 1.04 & 0.01 & 3.30  \\
HIP 44997 & 3.88 & 0.92 & 1.18 & 0.14 & -0.023 & 0.005 & 5731 & 5 & 4.47 & 0.02 & 0.95 & 0.01 & 2.93  \\
HIP 49756 & 4.62 & 0.57 & 0.73 & 0.16 & 0.043 & 0.004 & 5795 & 4 & 4.42 & 0.01 & 1.01 & 0.01 & 3.25  \\
HIP 54102* & 1.11 & 0.70 & 1.73 & 0.13 & -0.014 & 0.007 & 5820 & 9 & 4.51 & 0.02 & 1.02 & 0.02 & 3.18  \\
HIP 54287 & 6.34 & 0.40 & 1.33 & 0.13 & 0.118 & 0.004 & 5727 & 4 & 4.36 & 0.01 & 1.01 & 0.01 & 3.12  \\
HIP 54582* & 7.28 & 0.31 & 0.65 & 0.14 & -0.080 & 0.005 & 5875 & 7 & 4.27 & 0.02 & 1.17 & 0.01 & 3.82  \\
HIP 62039* & 6.73 & 0.44 & 1.89 & 0.12 & 0.088 & 0.005 & 5753 & 6 & 4.35 & 0.02 & 1.05 & 0.01 & 3.23  \\
HIP 64150* & 6.41 & 0.66 & 1.98 & 0.14 & 0.030 & 0.007 & 5747 & 6 & 4.39 & 0.02 & 1.00 & 0.02 & 3.13  \\
HIP 64673* & 5.22 & 0.55 & 1.86 & 0.17 & -0.030 & 0.007 & 5918 & 8 & 4.35 & 0.02 & 1.21 & 0.02 & 3.84  \\
HIP 64713 & 4.26 & 1.10 & 1.95 & 0.12 & -0.067 & 0.007 & 5767 & 8 & 4.46 & 0.02 & 1.00 & 0.02 & 3.08  \\
HIP 65708 & 9.41 & 0.28 & 1.35 & 0.15 & -0.066 & 0.006 & 5755 & 6 & 4.25 & 0.02 & 1.09 & 0.01 & 3.42  \\
HIP 67620* & 7.18 & 1.08 & 2.77 & 0.11 & -0.018 & 0.009 & 5670 & 9 & 4.41 & 0.03 & 1.01 & 0.03 & 2.83  \\
HIP 68468 & 5.90 & 0.40 & 1.92 & 0.13 & 0.065 & 0.007 & 5857 & 6 & 4.32 & 0.02 & 1.13 & 0.01 & 3.66  \\
HIP 69645$^\dagger$ & 5.27 & 0.85 & 2.05 & 0.11 & -0.045 & 0.006 & 5743 & 6 & 4.44 & 0.02 & 0.99 & 0.02 & 3.03  \\
HIP 72043* & 6.42 & 0.47 & 1.39 & 0.14 & -0.034 & 0.007 & 5842 & 8 & 4.35 & 0.02 & 1.12 & 0.02 & 3.55  \\
HIP 73241* & 9.38 & 0.35 & 2.08 & 0.11 & 0.082 & 0.007 & 5669 & 8 & 4.27 & 0.02 & 1.01 & 0.02 & 3.08  \\
HIP 73815 & 6.57 & 0.46 & 1.42 & 0.13 & 0.004 & 0.005 & 5788 & 6 & 4.37 & 0.02 & 1.05 & 0.01 & 3.32  \\
HIP 74389 & 1.01 & 0.48 & 1.09 & 0.19 & 0.077 & 0.004 & 5844 & 5 & 4.49 & 0.01 & 1.07 & 0.01 & 3.31  \\
HIP 74432**$^\dagger$ & 9.77 & 0.31 & 1.66 & 0.12 & 0.037 & 0.007 & 5684 & 8 & 4.25 & 0.02 & 1.09 & 0.02 & 3.17  \\
HIP 76114 & 6.15 & 0.82 & 1.30 & 0.13 & -0.037 & 0.006 & 5733 & 6 & 4.42 & 0.02 & 0.97 & 0.02 & 3.03  \\
HIP 77052** & 3.67 & 0.91 & 1.58 & 0.13 & 0.036 & 0.006 & 5683 & 5 & 4.48 & 0.02 & 0.96 & 0.02 & 2.75  \\
HIP 77883$^\dagger$ & 7.24 & 0.68 & 1.95 & 0.12 & -0.006 & 0.006 & 5690 & 6 & 4.40 & 0.02 & 0.99 & 0.02 & 2.92  \\
HIP 79578* & 2.17 & 0.78 & 1.74 & 0.12 & 0.057 & 0.005 & 5820 & 5 & 4.47 & 0.01 & 1.04 & 0.01 & 3.25  \\
HIP 79672$^\dagger$ & 3.09 & 0.39 & 2.08 & 0.11 & 0.056 & 0.003 & 5814 & 3 & 4.45 & 0.01 & 1.02 & 0.01 & 3.27  \\
HIP 79715 & 6.47 & 0.46 & 0.64 & 0.14 & -0.041 & 0.005 & 5803 & 6 & 4.38 & 0.02 & 1.09 & 0.01 & 3.35  \\
HIP 81746* & 7.53 & 0.58 & 1.43 & 0.12 & -0.086 & 0.004 & 5715 & 5 & 4.40 & 0.02 & 0.99 & 0.01 & 3.00  \\
HIP 83276** & 7.54 & 0.27 & 0.50 & 0.10 & -0.089 & 0.006 & 5885 & 8 & 4.22 & 0.02 & 1.23 & 0.01 & 3.95  \\
HIP 85042 & 6.66 & 0.62 & 1.64 & 0.14 & 0.015 & 0.004 & 5694 & 5 & 4.41 & 0.02 & 1.00 & 0.01 & 2.91  \\
HIP 87769* & 5.15 & 0.69 & 2.05 & 0.13 & 0.000 & 0.006 & 5807 & 6 & 4.40 & 0.02 & 1.05 & 0.01 & 3.33  \\
HIP 89650 & 3.82 & 0.76 & 1.67 & 0.14 & 0.000 & 0.005 & 5841 & 5 & 4.44 & 0.02 & 1.08 & 0.01 & 3.39  \\
HIP 95962** & 3.82 & 0.78 & 1.41 & 0.13 & 0.023 & 0.005 & 5806 & 5 & 4.44 & 0.02 & 1.04 & 0.01 & 3.26  \\
HIP 96160$^\dagger$ & 2.17 & 0.78 & 2.09 & 0.11 & -0.053 & 0.007 & 5781 & 8 & 4.50 & 0.02 & 0.96 & 0.02 & 3.06  \\
HIP 101905 & 1.59 & 0.69 & 3.06 & 0.11 & 0.057 & 0.006 & 5890 & 6 & 4.47 & 0.02 & 1.07 & 0.02 & 3.52  \\
HIP 102040 & 2.42 & 0.91 & 1.74 & 0.12 & -0.093 & 0.006 & 5838 & 6 & 4.48 & 0.02 & 1.05 & 0.02 & 3.30  \\
HIP 102152$^\dagger$ & 6.92 & 0.69 & 1.78 & 0.12 & -0.020 & 0.005 & 5718 & 5 & 4.40 & 0.02 & 0.95 & 0.01 & 3.01  \\
HIP 103983* & 2.08 & 0.86 & 3.38 & 0.10 & -0.048 & 0.008 & 5752 & 10 & 4.51 & 0.02 & 0.96 & 0.02 & 2.93  \\
HIP 104045$^\dagger$ & 2.29 & 0.83 & 2.09 & 0.11 & 0.045 & 0.005 & 5831 & 6 & 4.47 & 0.02 & 1.00 & 0.01 & 3.29  \\
HIP 105184 & 0.60 & 0.45 & 2.64 & 0.11 & -0.002 & 0.009 & 5833 & 11 & 4.50 & 0.02 & 0.99 & 0.02 & 3.25  \\
HIP 108158 & 8.36 & 0.48 & 1.20 & 0.15 & 0.067 & 0.008 & 5687 & 7 & 4.34 & 0.02 & 0.97 & 0.02 & 3.02  \\
HIP 108468 & 7.56 & 0.40 & 0.91 & 0.19 & -0.111 & 0.006 & 5829 & 7 & 4.33 & 0.02 & 1.16 & 0.01 & 3.54  \\
HIP 109821 & 9.30 & 0.39 & 0.82 & 0.17 & -0.115 & 0.005 & 5746 & 7 & 4.31 & 0.02 & 1.06 & 0.01 & 3.28  \\
HIP 114615 & 1.05 & 0.71 & 2.39 & 0.11 & -0.077 & 0.008 & 5816 & 9 & 4.52 & 0.02 & 1.04 & 0.02 & 3.15  \\
HIP 115577 & 9.50 & 0.34 & 1.32 & 0.15 & 0.036 & 0.008 & 5699 & 9 & 4.25 & 0.03 & 1.12 & 0.02 & 3.22  \\
HIP 116906$^\dagger$ & 6.46 & 0.44 & 1.74 & 0.12 & 0.010 & 0.005 & 5792 & 6 & 4.37 & 0.02 & 1.05 & 0.01 & 3.33  \\
HIP 117367 & 5.94 & 0.40 & 1.17 & 0.14 & 0.044 & 0.007 & 5871 & 8 & 4.32 & 0.02 & 1.15 & 0.02 & 3.72  \\
HIP 118115 & 7.79 & 0.32 & 0.89 & 0.19 & -0.017 & 0.006 & 5808 & 7 & 4.28 & 0.02 & 1.12 & 0.01 & 3.55  \\
Sun$^\dagger$ & 4.56 & \ldots & 2.04 & 0.12 & 0.000 & \ldots & 5777 & \ldots & 4.44 & \ldots & 1.00 & \ldots & 3.20

\label{params}
\end{longtable}
\end{center}
\tablefoot{* Spectroscopic binary star; ** visual binary star; $^\dagger$ \textit{selected sample} stars; $v\rm_{t}$ are the microturbulence velocities. $v_\mathrm{macro}$ are inferred from the scaling Eq. \ref{vmacro_eq}.}

\end{document}